\documentstyle[12pt]{article}

\def\be{\begin{equation}}
\def\ee{\end{equation}}
\def\bea{\begin{eqnarray}}
\def\eea{\end{eqnarray}}
\def\ba{\begin{array}}
\def\ea{\end{array}}

       \def\C{\rm {I\kern-.520em C}}
\def\s{\sum}

\def\t{\theta}
\def\d{\partial}
\def\de{\delta}
\def\f{\frac}
\def\a{\alpha}

\def\hs{\hspace}
\def\ff{\f{1}{2}}
\def\cp{{\cal P}}
\def\ca{{\cal A}}
\def\os1{$osp(2|1)$}

\begin{document}

\begin{titlepage}
\vspace{-10mm}

\begin{flushright}
          {\it   hep-th/9706164}\\
             IPM-97-208\\
             March 1997\\ 
\end{flushright} 
\begin{center}
\begin{large}
             {\bf Exactly and Quasi-Exactly Solvable Models\\on the Basis of
             \os1 }
\end{large}

\
\

{\bf A. Shafiekhani${}^\dagger$
\footnote{E-mail:
ashafie@theory.ipm.ac.ir}
and M. Khorrami${}^{\dagger *}$
\footnote{\hspace{1.2 cm}
mamwad@netware2.ipm.ac.ir}
}

\vspace{12pt} {\it
\vspace{8pt}
${^\dagger}$Institute for Studies in Theoretical Physics and Mathematics\\
             P.O.Box: 5531, Tehran, 19395, Iran,\\
\vspace{0.3cm}

${^*}$ Department of Physics, Tehran University,\\
             North-Kargar Ave. Tehran, Iran.\\
                             and\\
Institute for Advanced Studies in Basic Sciences,\\
             P.O.Box: 159, Gava Zang, Zanjan 45195, Iran. \\
}

\vspace{0.3cm}

\

\end{center}
\abstract{
The exactly and quasi-exactly solvable problems for
spin one-half in one dimension on the basis of a hidden dynamical symmetry
algebra of Hamiltonian are discussed. We take the supergroup, $OSP(2|1)$,
as such a symmetry. A number of exactly solvable examples are considered
and their spectrum are evaluated explicitly. Also, a class of quasi-exactly
solvable problems on the basis of such a symmetry has been obtained.
\
\vfill
}

\end{titlepage}
{\section{Introduction} }
In the last few years, a new class of quantum mechanical problems has
been investigated. This is the class of quasi-exactly solvable problems [1-6],
which in some sense is an intermediate  class
between problems for which the spectrum can be found exactly,
analytically or algebraically, and those which can not be solved.
A hidden symmetry group with a finite dimensional space of wave functions is
responsible for the existence of quasi-exactly solvable problems
\cite{t1,shif}.
Such a symmetry
is inherent to any exactly and quasi-exactly solvable problem. In the
exactly solvable problems case, the symmetry is, in some sense, complete,
and the Hamiltonian
can be diagonalized. In the case of quasi-exactly solvable models, the
Hamiltonian is only block diagonalized, and there is a finite block which
can be diagonalized. Therefore only part of the spectrum can be found. In
both cases the Hamiltonian can be reduced to at most quadratic combinations of
generators of a relevant symmetry group with finite dimensional representations.
Such Hamiltonians can be written in standard form by choosing differential
realizations of generators. Differential realizations of generators are
representations of generators in a finite dimensional subspace of analytical
functions \cite{az}. Hence, one will be able to compute the spectrum
by using these realizations.
There have been attempts to do so for spinless \cite{t1,shif,sht} and
spin one-half [5, 8-11] particles. In this paper the supergroup $OSP(2|1)$
has been taken as symmetry group of a spin one-half
particle.

Differential realizations of the generators of this group contains
beside $z$ and $\f{\d}{\d z}$, $\t$ and $\f{\d}{\d\t}$, where $\t$ is a
a Grassmann variable. Therefore the
Hamiltonians are function of
$z$, $\f{\d}{\d z}$, $\t$, and $\f{\d}{\d\t}$. To get the standard
form of the Hamiltonian, we use proper change of variable from $z$ to $x$,
take $x$ as the position variable, and put
\be
\t:=\left (\ba{cc} 0&0\cr 1&0\ea\right ),\hs{5 mm}
\f{\d}{\d\t}:=\left (\ba{cc} 0&1\cr 0&0\ea\right ).
\ee
Hence, we will have standard form of Hamiltonian with spin interaction terms.

Our aim in this paper is to give a new class of exactly and quasi-exactly
solvable problems. The structure of this paper is as follows: In section 2,
we construct general exactly solvable Hamiltonians for spin one-half
particles on the basis of the supergroup $OSP(2|1)$, and find the
corresponding spectrum.
In section 3 we generalise the Hamiltonians given
in section 2 to a class of quasi-exactly solvable
problems for spin one-half particles.

\section{Exactly Solvable Models}
One of the major symmetry group candidates for spin one-half particles is the
supergroup with rank one,
$OSP(2|1)$.
Such a group has three even and two odd generators.
The differential realization of the algebra \os1 is obtained in \cite{sht,wa}:
\be\ba{ccc}\label{gen}
J_-=\d &=:& J_{-1}\cr
J_+=-z^2\d+2hz-z\t\de&=:& J_{+1}\cr
J=2z\d-2h+\t\de&=:& J_0\cr
G_-=\de+\t\d&=:& J_{-\ff}\cr
G_+=z\de+\t z\d-2h\t&=:& J_{+\ff}
\ea\ee
where $\{J_\pm,J_0\}$ are even and $\{G_\pm\}$ are odd generators,
$\t$ is a Grassmanian variable, $h$ is half of the spin of the representation,
and $\d=\f{\d}{\d_z}$, $\de=\f{\d}{\d_\t}$.

One can consider
this, as the representation
of \os1 on the super-sub-space of analytic functions spanned by the monomials,
\be
{\cal F}_h:=\{z^p\t^q\mid p=0,1,\cdots, 2h,\;q=0,1\}.
\ee
These generators satisfy the (anti)commutation relations
\be\ba{ccc}
[J_+,J_-]=J & \{G_\pm,G_\pm\}=\mp 2J_\pm & \{G_+,G_-\}=J\cr
[J,J_\pm]=\pm 2J_\pm & [J,G_\pm]=\pm G_\pm & [J_+,G_-]=G_+\cr
[J_+,G_+]=0 & [J_-,G_+]=G_- & [J_-,G_-]=0
\ea\ee
The Casimir of this realizations is
\be\label{cas}
J^2_0+2\{J_-,J_+\}+[G_-,G_+]=2h(2h+1)
\ee
To write a solvable Hamiltonian, we note that the generator $J_\a$ ( one can
call $\a$ the spin) acting on $z^p\t^q$, changes the index $p+\f{q}{2}$ to
$p+\f{q}{2}+\a$. Moreover, the set of generators (\ref{gen}) acting on states
with $p+\f{q}{2}\ge 0$, give rise to states with $p+\f{q}{2}\ge 0$. Hence, 
one can regard the state $z^{p=0}\t^{q=0}$ as the ground state. Now, any 
combination of
operators with total $\a$ not greater than zero has a matrix representation
which is upper triangular, where the ordering of the states $z^p\t^q$ is
according to $p+\f{q}{2}$.
The eigenvalues of such a (n infinite dimensional)
matrix are its diagonal elements and the eigenvectors can be obtained 
recursively.

Hence, to write a solvable Hamiltonian it is sufficient to add terms of the
type $J_{\a_1}\cdots J_{\a_n}$, where $\s\a_i\le 0$. As we want to convert
such a Hamiltonian to a Schr\"odinger one, that is one with a second order
derivative and a potential, we must keep only terms at most quadratic in
the generators.
Moreover,  (\ref{cas}) shows that not all second order generators are
independent.
We can list all up to second order independent spin-non-positive generators.
\be\ba{lcl}\label{snpg}
spin\; -2    &:& A:=\d^2=J^2_-\cr
spin\; -3/2  &:& \d\de+\d^2\t=\ff\{J_-,G_+\}\cr
spin\; -1    &:& \left\{\ba{l}{B:=\d=J_-}\cr
                   {C:=4z\d^2+2\d\t\de=\{J,J_-\}+(4h-2)J_-}\ea\right.\cr
spin\; -1/2  &:& \left\{\ba{l}
                   \de=\f{1}{2(1+4h)}[(1+4h)G_-+2\{J_-G_+\}-\{J,G_-\}]\cr
                   \t\de=\f{1}{2(1+4h)}[(1+4h)G_--2\{J_-,G_+\}+\{J,G_-\}]\cr
                   z\d\de+z\d^2\t=\f{2h-1}{2}G_-+\f{1}{2(1+4h)}[\{J_-,G_+\}
                   +2h\{J,G_-\}]
                   \ea\right.\cr
spin\;\; 0    &:&  \left\{\ba{l}
                   D:= z\d=\ff(J+2h)-\f{1}{2(1+4h)}([G_-,G_+]+2h)\cr
                   E:= \t\de=\f{1}{1+4h}([G_-,G_+]+2h)\cr
            \ba{ll}F:= z^2\d^2+z\d\t\de=&\f{1}{4}[(J+2h)^2-2J-4h]\cr
                                      &+\f{1}{4(1+4h)}([G_-,G_+]+2h)\ea
                  \ea\right.
\ea\ee
Another thing to be considered is that we want to use this combination and then
make a change of variable $z\rightarrow x$ and a quasi gauge transformation to
bring the Hamiltonian to a standard form.
Now, as beside $z$ there exist $\t$ and $\de$ in Hamiltonian, this change of
variable depends on $\t$ and $\de$, which shows that $x$ no longer commutes
with $\t$ and $\de$. However, if only the combination $\t\de$ appears in
the Hamiltonian, $x$ will contain only $z$ and $\t\de$, and the final
Hamiltonian will consist of just $x$
and $\t\de$. In this case, although $x$ does not commute with $\t$
and $\de$, it does commute with $\t\de$ and the Hamiltonian has its
standard form: having a space degree of freedom and some internal
degree with a combination of this parameter
which commutes with the space degree of freedom.

So, the Hamiltonian which we consider is of the form
\be\label{exh1}
{\tilde H}= aA+bB+cC+dD+eE+fF.
\ee
This Hamiltonian acts on the space spanned by the basis $\{z^p\t^q\}$, where
$p$ is nonnegative and integer, and $q=0,1$. ${\cal F}_h$ is a subspace of
this space.

As it is shown below, one can use the variable $x$, instead of $z$ and make
a quasi gauge transformation to make $H$ in the form
\be
H=-\f{\hbar^2}{2m}\f{d^2}{dx^2}+V(x,q).
\ee
Here one can interpret $x$ as the space coordinate and $q$ as the matrix
$\left (\ba{cc} 1&0\cr 0&0\ea\right )$. Then $H$ is a Hamiltonian of an
spin $1/2$ system, and obviously Hermitian with respect to the inner product
$<f,g>:=\int f^\dagger(x)g(x) dx$, where $f$ and $g$ are functions
${\cal R}\rightarrow \C^2$.
{\subsection{Explicit Examples} }
The Hamiltonian (\ref{exh1}) can be written in the form
\be\label{exh2}
{\tilde H}= (fz^2+4cz+a)\f{\d^2}{\d z^2}+[(b+2cq)+(d+fq)z]\f{\d}{\d z}+eq
\ee
where
\be
q:=\t\de .
\ee
The spectrum of this Hamiltonian is simply the diagonal elements of it in
$z^p\t^q$ representation, that is
\be
E_{p,q}=fp[(p-1)+q]+dp+eq.
\ee
Now consider the following cases.\\
{\bf{i)} $c=f=0\neq a$}\\
In this case,  (\ref{exh2}) becomes
\be
{\tilde H}=a[\f{\d}{\d z}+\f{b+dz}{2a}]^2-\f{(b+dz)^2}{4a}-\f{\d}{2}
+eq
\ee
and, by a simple change of variable and a pseudo-gauge
transformation \cite{shif,sht},
to eliminate the first order derivative,
\be\label{ex1}
H=a\f{\d^2}{\d x^2}-\f{d^2}{4a}x^2-\f{d}{2}+eq.
\ee
Using
\be
\left\{\ba{l} a=-\f{\hbar}{2m}\cr d=\hbar w \ea\right.
\ee
it is seen that (\ref{ex1}) is the Hamiltonian of a Harmonic oscillator which
is also confined in a magnetic field. The spectrum of (\ref{ex1}) is
\be
E_{p,q}=dp+eq
\ee
{\bf ii)} $f=0\neq c$\\
The Hamiltonian becomes
\be
{\tilde H}=(a+4cz)\f{\d^2}{\d z^2}+(b+dz)\f{\d}{\d z}+2cq\f{\d}{\d z}+eq.
\ee
Using
\be
x=\sqrt{z+\f{a}{4c}}
\ee
and a proper pseudo-gauge transformation, we arrive at
\be\label{ex2}
H=c\f{\d^2}{\d x^2}-\f{d^2}{16c}x^2-\f{d}{4}-\f{c\zeta(1+\zeta)}{x^2}
+\f{d\zeta}{2}+eq
\ee
where
\be
\zeta :=\f{2c-2cq-b+ba/(4c)}{4c}.
\ee
Now, setting
\be
\left\{\ba{l} c=-\f{\hbar^2}{2m}\cr d=2\hbar \omega\ea\right.
\ee
it is seen that $H$ is very similar to the radial part of the
Hamiltonian of an
isotropic Harmonic  oscillator in a magnetic field. There are,
however, nontrivial interactions; the terms $\f{q}{x^2}$ and
$\f{q^2}{x^2}$, which represent spin-nonuniform magnetic field interactions.
The spectrum of this Hamiltonian is
\be
E_{p,q}=dp+eq.
\ee
{\bf iii)}$f\neq 0$\\
By a suitable shift in $z$, (\ref{exh2}) becomes
\be
{\tilde H}=(a+fz^2)\f{\d^2}{\d z^2}+[b+(d+fq)z]\f{\d}{\d z}+eq.
\ee
Then, using
\be
z=\sqrt{\f{a}{f}}\sinh{x},
\ee
and a proper pseudo-gauge transformation,
\be\label{ex3}\ba{lcl}
H&=&-f[-\f{\d^2}{\d x^2}+\f{(q-1+d/f)^2}{4}\tanh^2{x}+(\f{b^2}{4af}
+\f{q-1+d/f}{2})\cosh^{-2}{x}\cr
&+&\f{b}{2\sqrt{af}}(2-q-\f{d}{f})
\f{\sinh{x}}{\cosh^2{x}}+eq.
\ea\ee
The spectrum of this is
\be
E_{p,q}=fp(p-1)+fpq+dp+eq.
\ee
Note that all of the Hamiltonians so obtained, are Hermitian.
{\section{Quasi-Exactly Solvable Problems}}
Up to now, we investigated only the exactly solvable problems.
To go beyond, and investigate
quasi-exactly solvable problems, we should also consider
the spin-positive generators.
\be\ba{lcl}\label{spg}
spin\;  2  &:& A':=z^4\d^2-2kz^3\d+2hkz^2+[2z^3\d-2kz^2]\t\de=J_+^2\cr
spin\; 3/2 &:& [-z^3\d+kz^2]\de+[-z^3\d^2+2kz^2-2hkz]\t=\ff\{J_+,G_+\}\cr
spin\; 1   &:& \left\{\ba{l}{B':=-z^2\d+2hz-z\t\de=J_+} \cr
                    {C':=-2z^3\d^2+2kz^2\d+[-3z^2\d+kz]\t\de=\ff\{J,J_+\}
                    +kJ_{+\ff}}
                    \ea\right.
\ea\ee
where $2h-1=k$. By the same reason of section 2, the Hamiltonian should consist
of just $z$ and $\t\de$, so final Hamiltonian which we consider for
quasi-exactly solvable problem is of the form
\be\label{qh}
{\tilde H}=aA+a'A'+bB+b'B'+cC+c'C'+dD+eE+fF.
\ee
The above Hamiltonian can be written in form of
\be\ba{lcl}
{\tilde H}&=&(a'z^4-2c'z^3+fz^2+4cz+a)\d^2+[2a'(q-k)z^3+((2k-3q)c'-b')z^2\cr
&+&(d+fq)z+(b+2cq)]\d+2a'k(h-q)z^2+[c'kq+b'(2h-q)]z+eq\cr
&:=& \cp_4(z)\d^2+\cp_3(z,h,q)\d+\cp_2(z,h,q).
\ea\ee
In the above expression for Hamiltonian $h$ plays an important role for
quasi-exactly solvable problems. The Hamiltonian matrix is in block-diagonal
form with two blocks: one is $2h+1$-dimensional and the other is
infinite-dimensional. The former block can be diagonalized, as it is finite
dimensional.

To bring the Hamiltonian to the standard form, the following change of
variable should be done
\be
x={\a}\int \f{dz}{\sqrt{\cp_4(z)}}.
\ee
Using the above change of variable the Hamiltonian will be
\be
{\tilde H}=\a\f{\d^2}{\d x^2}+2\a\ca (x,q,h)\f{\d}{\d x}+\cp_2(x,q,h)
\ee
where
\be
\ca (x,q,h)=\f{1}{4\a} [
\f{2\cp_3(z,q,h)-\cp_4'(z)}{\sqrt{\cp_4(z)}}]
\mid_{z=z(x)},\hs{0.5 cm} \ca=\f{\d a(x)}{\d x}.
\ee
In the above expression $a(x)$ is pseudo-gauge, and potential is
\be
V(x,q,h)=\cp_2(x,q,h)-\a(\f{\d\ca}{\d x}-\ca^2).
\ee
Putting all together we have
\be
{\tilde H}{\hat \psi}=E{\hat \psi}
\ee
where
\be
\psi=e^{-a(x)}{\hat \psi},\hs{0.5 cm} \{{\hat \psi}\}={\cal F}_h\cup\{\psi_{2h
     +2},
\psi_{2h+2}\t,\cdots\}.
\ee
In this way, one gets a Hamiltonian which is a similarity transformation of
a block diagonal operator, having a finite block (corresponding to ${\cal F}_h$
).
So, the eigenvalues (and eigenvectors) corresponding to this space can be 
obtained.

\end{document}